# Audio Interval Retrieval using Convolutional Neural Networks

Ievgeniia Kuzminykh[1][0000-0001-6917-4234], Dan Shevchuk[2], Stavros Shiaeles[3][0000-0003-3866-0672], and Bogdan Ghita[4][0000-0002-1788-547X],

[1] King's College London, Strand, London, WC2R 2LS, UK
`ievgeniia.kuzminykh@kcl.ac.uk`
[2] Blekinge Institute of Technology, Karlskrona, Sweden Campus Grasvik, SE- 371 41
[3] University of Portsmouth, Portsmouth, PO1 3RR, UK
`stavros.shiaeles@port.ac.uk`
[4] University of Plymouth, Drake Circus, Plymouth, PL4 8AA UK
`bogdan.ghita@plymouth.ac.uk`

**Abstract.** Modern streaming services are increasingly labeling videos based on their visual or audio content. This typically augments the use of technologies such as AI and ML by allowing to use natural speech for searching by keywords and video descriptions. Prior research has successfully provided a number of solutions for speech to text, in the case of a human speech, but this article aims to investigate possible solutions to retrieve sound events based on a natural language query, and estimate how effective and accurate they are. In this study, we specifically focus on the YamNet, AlexNet, and ResNet-50 pre-trained models to automatically classify audio samples using their respective melspectrograms into a number of predefined classes. The predefined classes can represent sounds associated with actions within a video fragment. Two tests are conducted to evaluate the performance of the models on two separate problems: audio classification and intervals retrieval based on a natural language query. Results show that the benchmarked models are comparable in terms of performance, with YamNet slightly outperforming the other two models. YamNet was able to classify single fixed-size audio samples with 92.7% accuracy and 68.75% precision while its average accuracy on intervals retrieval was 71.62% and precision was 41.95%. The investigated method may be embedded into an automated event marking architecture for streaming services.

## 1 Introduction

Natural language search is becoming more and more popular among users, with an increasing number of mobile and embedded devices providing the feature either included in the operating system or as part of a streaming services. This is evidenced by the sale of devices that support this feature. Due to this increased market interest, recent years have witnessed a significant amount of research on factors influencing engine rankings during voice search [1]. In early 2019, Amazon released device sales data for which it sold over 100 million devices that support the voice assistant applications [2]. Demand for these devices exceeded supply by several times, therefore, some models are currently impossible to purchase. Other studies from WARC, Dynata and PwC also confirms the popularity of voice searches [3,4]. According to their research, 60% of the mobile device users actively use voice search, and mobile advertising reached 62.4% of the user population. More than half of users (53%) who make a request from a smartphone are looking for information about a product, 50% find answers to questions of interest, 42% use navigation and 41% look for information about brands.

The usability and convenience of voice search subsequently led to an expansion of the scope of this function, with searching for a video fragment by description being a very popular choice. Most of the time people can describe verbally what they are looking for but, given the lack of metadata associated with the timeline and content of the video, this option is currently mostly unavailable. however, currently there is no publicly available efficient way of searching through multimedia content to identify and label an audio event or sound.

Numerous studies have been conducted in the area of audio recognition over the past few decades [5-9]. Most of the traditional approaches require handcrafted features to provide metadata. This situation improved significantly with the use of modern deep neural network architectures that rely on convolutional layers [10]. While up until recently it was possible to achieve a reasonable level of feature extraction from images, now it is also possible to effectively recognise patterns in videos using deep neural networks. The most recent research [6] achieves a very reliable result. However, it aggregates features from 3 sources: individual frames, an array of frames at once (continuous snippets) and subtitles. Its success can further be improved by extracting the features from the soundtrack, which the paper does not do. To exemplify one should consider a scene where the actor stays silent and only the sound of an airplane can be heard in the background, but the aircraft is not visible in the scene. Querying video or subtitles for "sound of a flying airplane while the actor is standing still" cannot yield any helpful result because the required input is not present either in the video or in the subtitles. Another example is a generic nature scene with an arbitrary song in the background. The user can remember the song but we would need to also search through the soundtrack to use it for query. Such examples demonstrate that this result can be improved by also including the soundtrack as one of the input parameters. In this context, the biggest challenge is the lack of technology to allow interval retrieval from sound data based on semantic input, in order to replicate and support the existing technology for video interval retrieval.

The aim of this study is to identify and investigate models capable of retrieving intervals from an audio sample based on a natural language query. To benchmark their efficiency, the models will be trained on the AudioSet [11] database that consists of an expanding ontology of 632 audio event classes and a collection of 2,084,320 human-labeled 10-second sound clips drawn from YouTube videos. The ontology is specified as a hierarchical graph of event categories, covering a wide range of human and animal sounds, musical instruments and genres, and common environmental sounds.

## 2　Related Works

There have been a few studies conducted in different areas that relate to the challenges of audio information retrieval based on semantic input. A significant subset of this body of research proposed a number of methods for querying an audio file by the semantic description of its class [12-16]. These can be considered partial solutions when compared to the aim of this study, which is not only to identify the class of the given audio but also to convert its semantic description into an abstraction that can be matched to the extracted features of the given audio file of an arbitrary length.

Aside from actual querying, research also aimed to provide a better, more distinctive input through data pre-processing, analyse the features of the audio data to be fed into the models, and identify the best approaches for one of the intermediate data segmentation and classification steps. From this line of research, the approach from Qazi et al. involved segregating the audio intervals of interest (such as music and speech) from noisy data [17]. This may provide a useful and necessary pre-processing step to clear up the audio input before starting to extract features that can be matched to the semantic description, thus improving the accuracy of the resulting model. Along the same direction, a number of studies [18-21] sought to investigate the most useful features of the audio data, which would subsequently allow them to analyse and retrieve information from it more accurately. Others research [22-25] redesigned the audio data classification process to allow searching for an audio file or to automatically label existing audio clips.

On the wider societal context, Pfeiffer et al. sought to automate the process of audio data analysis with different applications such as music indexing and retrieval as well as violence detection in the sound track of videos [26]. Foote J. has studied the process of audio information retrieval in general to give insights into the process that can be valuable for any kind of further research in this area [27].

In terms of the core method used, Shawn Hershey et al. have compared the VGG, AlexNet, ResNet-50, Inception V3, and Fully-Connected architectures on their performance on audio data. Based on a 17 million samples dataset, the study identified ResNet-50 as the best performing model [10]. However, the analysis focused exclusively on the performance of the models on sound classification. The evaluation can be extended by using the models for audio interval retrieval based on the natural language query and comparing the models' performance in terms of accuracy, precision etc.

All of the studies that are mentioned above provide either a partial solution to the problem of audio-interval retrieval based on a semantic description, which we are seeking to solve with this study, or provide a useful information that will help address this thesis' aim faster and more accurately.

Though there are reliable methods of event retrieval based on semantic input, they are all based solely on the visual data, like individual frames or motion features. To the best of our knowledge, none of the related work papers we surveyed employ the soundtrack for increasing the accuracy of the interval retrieval. Therefore, studying the best performing methods for audio-based interval retrieval would improve the existing technology of content-based video and audio classification and searching in a unified approach and would enable further analysis of audio data.

## 3 Method

This study investigates the effectiveness of deep learning models within the convolutional neural networks of sound classification. To benchmark the performance of such algorithms, the experiment involved three models: AlexNet, ResNet-50 and YamNet. Both AlexNet and ResNet-50 are known to be well-suited for image classification, as highlighted in the previous section. In order to exploit their ability to discriminate visual input, the input audio data used for training and testing the models was converted into melspectrograms, providing them with the necessary input [10, 18]. YamNet model is an already pretrained model provided by TensorFlow team. YamNet takes in a waveform of the given sound data sample and predicts the probability of each of its 521 classes; it was chosen for the study due to the fact that it is pretrained on the same dataset used for this study.

### 3.1 Performance Metrics

In order to evaluate the performance of the models and comparatively benchmark them, we followed the procedure outlined in previous studies [28, 29] and calculated Accuracy, Precision, F-1 score, MCC, and ROC AUC.

The metrics operate with true and false positives and negatives measurements to be calculated. The nature of true and false positives and negatives is explained in Table 1.

**Table 1.** True and False Negatives and Positives.

|  | NO (Prediction) | YES (Prediction) |
|---|---|---|
| NO (Actual) | True Negative (TN) | False Positive (FP) |
| YES (Actual) | False Negative (FN) | True Positive (TP) |

The first metric used to compare the models is the receiver operating characteristic area under curve (ROC AUC). The ROC curves are constructed by plotting the false positive rate (FPR) on the x-axis against the true positive rate (TPR) on the y-axis at various threshold settings. After calculating the ROC curves, the area under the curve (AUC), is computed for each ROC curve. The area under the curve is equal to the probability that a classifier will rank a randomly chosen positive instance higher than a randomly chosen negative one (assuming 'positive' ranks higher than 'negative') [29].

The Matthews correlation coefficient (MCC) is used in machine learning as a measure of the quality of binary (two-class) classifications. The metric is generally regarded as a balanced measure that can be used even if the classes are of different sizes [29].

$$\text{MCC} = \frac{\text{TP}\cdot\text{TN}-\text{FP}\cdot\text{FN}}{\sqrt{(\text{TP}+\text{FP})\cdot(\text{TP}+\text{FN})\cdot(\text{TN}+\text{FP})\cdot(\text{TN}+\text{FN})}} \quad (1)$$

Accuracy is closeness of the measurements to a specific reference value.

$$Accuracy = \frac{TP + TN}{TN + FP + TP + FN}. \tag{2}$$

The precision of a model is defined as the number of True Positives divided by the sum of all positive predictions.

$$Precision = \frac{TP}{TP + FP}. \tag{3}$$

A F measure (or sometimes F1 score) is a statistic that tries to combine the concerns of precision and recall in the same metric. F measure is calculated using the following Equation:

$$F1 = \frac{2 \cdot (recall \cdot precision)}{recall + precision}. \tag{4}$$

The above parameters provide a full statistical overview of the accuracy of the method, beyond its ability to discriminate individual events.

### 3.2 Data

The dataset used for evaluation is based on the AudioSet [11] database that consists of an expanding ontology of 527 audio event classes and a collection of 2,084,320 human-labelled 10-second sound clips drawn from YouTube videos. The ontology is specified as a hierarchical graph of event categories, covering a wide range of human and animal sounds, musical instruments and genres, and common everyday environmental sounds. Every video has on average 5 event labels. A combination of metadata (title, description, comments, etc.), context, and image content for each video were used to generate the labels automatically. It is worth noting that, in the dataset, the labels apply to the entire video and the labels are generated automatically, hence they are not 100% accurate [10].

The AudioSet database contains links to Youtube videos, labels and time frames for the correct segments from distinct videos. But not all links are available in the moment, therefore, a subset of 16902 samples was used for evaluation. The resulting number of items from the training dataset was 19614 items. We have decided to use only 6 out of 527 audio classes due to simplification of the experiment, and due to time and computational constraints. The selected classes are: "Rapping", "Cheering", "Gunshot, gunfire", "Radio", "Cat" and "Helicopter".

The reduced dataset is balanced. The downloaded evaluation and training datasets were merged and only audio samples that have one of the selected classes as labels were picked for the experiment as the final training and evaluation dataset. For the purpose of obtaining a higher degree of accuracy, all the entries from both training and evaluation dataset were merged and then randomly split 80/20 for as the training and testing subsets.

Each of 6 classes has approximately 120 audio samples linked to them in the training data had been picked and then another 11266 augmentations had been generated so that each class is represented by approximately 2000 samples in the training dataset. Totally, 12000 were used for training, of which 734 are 10 seconds raw samples in wave format.

After the last layers of AlexNet and ResNet-50 models were fitted, an additional classifier, Random Forest, was trained for each model, including YamNet.

### 3.3 Interval retrieval

To understand the performance of the models on a fixed-size length audio samples, 184 audio samples out of 16902 from the evaluation dataset of AudioSet [11] were used. The selection criteria for this data was that an audio sample contains keyword(s) belonging to one or more tested classes. Each sample gets split into fixed-size non-overlapping patches and fed into the classifiers. The evaluation dataset does not contain any augmentations, only raw original data. For each model, the output was fed into the instance of the Random Forest classifier trained for that specific model and the final output was evaluated against the target. The type of classification in this experiment is also binary multilabel classification.

In order to evaluate the performance of the models, a new audio file was created. The file is 41 seconds long and it contains samples from 3 audios, each of which is present in the evaluation dataset for the classification problem. The audio file is constructed in a way presented on Fig.1 that has distinct intervals of sounds that are easily recognizable by a human listener and has intervals without any sound as well. This melspectrogram is fed into all the models that have been already trained with the data. Since the evaluation audio file was handcrafted, it is labeled with a high degree of accuracy. The predicted intervals were compared against the handmade labels.

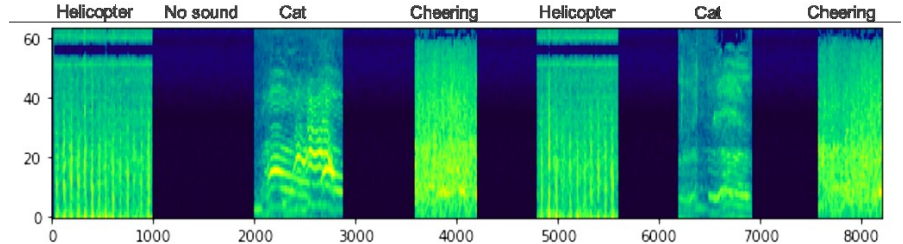

**Fig. 1.** A melspectrogram obtained by transforming the handcrafted evaluation audio file.

Comparing the predicted intervals with the handmade labels involves creating binary arrays for both predicted intervals and for hand-made ones. The binary array is an output in the binary multilabel classification, whereby the interval is retrieved using the user-provided keyword that should match one of the classes in the fixed term (in this case a fixed size of 0.01 seconds). Each patch is an array of length 6 but in the case of our audio file from Fig.1 three out of six elements in the array will be 0. Each element of the output arrays is a float number in range [0, 1] and represents the probability of each of the predicted classes.

### 3.4 Models

**YamNet.** It is a deep neural network that, following training predicts 521 audio event classes based on the AudioSet-YouTube corpus, and employing the Mobilenet_v1 depthwise-separable convolution architecture [30]. Six classes out of 527 were omitted based on the recommendations from reviewers [10].

The MobileNet structure is built on depthwise separable convolutions except for the first layer which is a full convolution. All layers are followed by a batchnorm and ReLU nonlinearity with the exception of the final fully connected layer, which has no nonlinearity and feeds into a softmax layer for classification. Down sampling is handled with strided convolution in the depthwise convolutions as well as in the first layer. A final average pooling reduces the spatial resolution to 1 before the fully connected layer. Counting depthwise and pointwise convolutions as separate layers, MobileNet has 28 layers.

As mentioned above, the model normally outputs 521 classes. However, for the purposes of the experiment, the output has been filtered and the array of 521 elements was replaced with a 6-element array, each of which represents the probability of each of the 6 classes picked for the study.

The output is then fed into another classifier - Random Forest. The final output is the one used for intervals retrieval since the model performs better together with the classifier than on its own.

**AlexNet.** It contains eight layers - five convolutional and three fully-connected [31]. The original model was modified for this task, with the last layer changed to output an array of 6 classes instead of the default 1000. We used the pretrained version of the model, which was trained to classify natural images. We decided to use the deep features of the model as well retrain it so that the model fits the problem better. The Random Forest classifier was also used on top of the output of AlexNet.

**ResNet-50.** The original ResNet-50 is a residual neural network which consists of 50 layers, most of which are convolutional [32]. The model was altered similarly to AlexNet, with the last layer set to output an array of 6 classes instead of the default 1000. The pretrained version of the model was used, which is pretrained to classify natural images. We decided to use deep features of the model as well retrain nine last layers. We also used a Random Forest classifier on top of the ResNet-50's output.

## 4 Results and Analysis

To understand the performance of the models on fixed-size audio samples, they were applied on the evaluation dataset. Table 2 shows the result value of the measured metrics in percentage for all three models. For comparison purposes, the YamNet model was implemented without the additional Random Forest classification in the experiment.

The best performing model overall is a combination of YamNet and Random Forest. However, comparing to the pure YamNet model, without Random Forest, the other models with additional classifier in form of Random Forest showed better results. To be more specific, the reason for the difference in performance lies in the fact that YamNet was trained on 2084320 human-labeled 10-second sound clips drawn from YouTube videos, while AlexNet and ResNet-50 were trained on 12000 samples, 11266 of which were automatically generated from the original 734.

**Table 2.** Models' performance on classifying audio samples.

|  | YamNet without RF | YamNet | AlexNet | ResNet-50 |
|---|---|---|---|---|
| ROC AUC | 62.5 | 92.5 | 87 | 88 |
| MCC | 24.9 | 81.53 | 65.8 | 67.6 |
| Accuracy | 79.16 | 92.7 | 88.7 | 88.85 |
| Precision | 37.5 | 68.75 | 61.74 | 63.11 |
| F1 score | 42.85 | 75.25 | 71.4 | 72.63 |

The next experiment was conducted in order to see the performance of the model on the classification of a single sample correlated with the performance on retrieving the intervals by querying the specific classes. As input the hand-crafted file was used. The performance of the models for three classes is presented on the Fig.2.

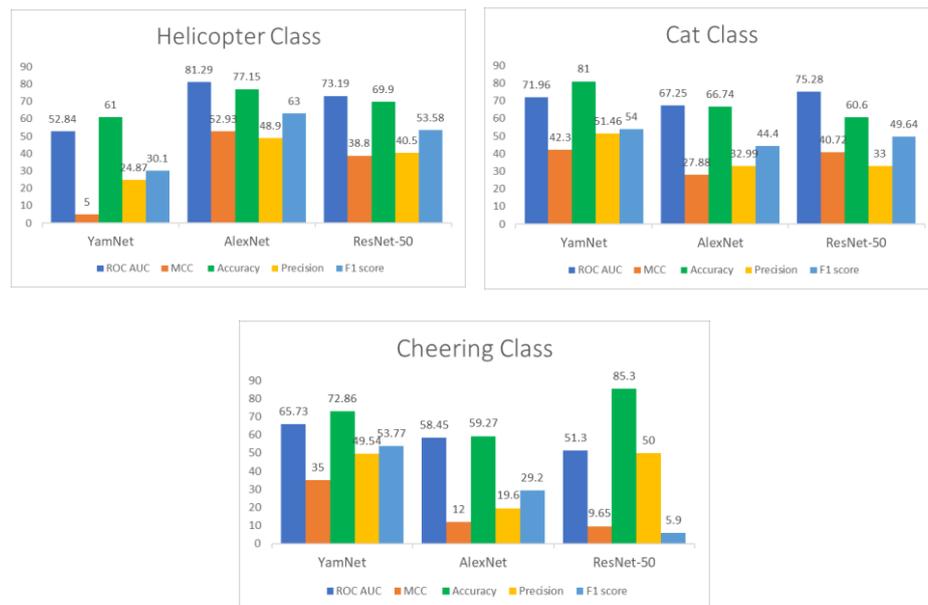

**Fig. 2.** Comparison of the metrics on the intervals retrieval by keywords.

The best performing model for class "Helicopter" was AlexNet which outperforms the other models on every single metric. The performance of AlexNet in this case is far from perfect, with an accuracy of 77.15% and a precision of 48.9%. In addition, the values of the other metrics suggest that it could be possible to achieve a reliable result with more training data. YamNet yielded a very low value for MCC even though the model had been trained on more examples than the others. This is possibly due to the fact that YamNet was trained to recognize 521 classes and its output was just filtered and the 6 classes were used to construct the final output while the other two models were fitted for these specific 6 classes.

The best performing model for class "Cat" was YamNet with accuracy of 81% and precision of 51.46%. AlexNet and ResNet-50 yielded a similar result, but ResNet-50 still performed better, yielding higher values for both ROC AUC (75.28%) and F1 score (49.64%).

The best performing model for class "Cheering" was YamNet, with accuracy of 72.86% and precision of 49.54%. ResNet-50 yields an even higher accuracy and precision, however, a low value of F1-score suggests that the high values of some of the other metrics are a result of the model not activating the queried class on a disproportionately large part of the audio sample rather than a result of good performance.

As we can see from Table 3, the aggregated values of the models' performance metrics are comparable. However, we know from examining the classes one by one that some of the values are too high on some classes and too low on the others just like in case of ResNet-50. Having examined the performance of the models on a single class case by case we can say that the best performing model overall is YamNet, yielding a mean accuracy of 71.62% and a mean precision of 41.95%. The model also yields stable values for most of the other metrics.

Table 3. Mean performance on three queried keywords.

|  | YamNet | AlexNet | ResNet-50 |
|---|---|---|---|
| ROC AUC | 67 | 69 | 66.59 |
| MCC | 27.4 | 30 | 29.72 |
| Accuracy | 71.62 | 67.72 | 71.93 |
| Precision | 41.95 | 33.83 | 41.16 |
| F1 score | 45.95 | 45.53 | 36.37 |

## 5 Discussion

In order to find the most suitable matching method to retrieve an audio of an arbitrary length via a natural language query input, two experiments have been conducted. Since the data that needs to be classified is audio and the features that were extracted from the data are melspectrograms, we have picked the models that perform well on image classification [10, 30]. The selected models are AlexNet and ResNet-50. All models have been pretrained to classify audio samples.

Two sets of tests were conducted to find the best performing model. The first test evaluated the performance of the models on the fixed-size audio samples classification problem. The best performing model turned out to be YamNet with accuracy of 92.7% and precision of 68.75%.

The purpose of the second test was to evaluate the performance on the intervals retrieval problem. The best performing model is still YamNet, but it is worth noting that the aggregated value of the metrics is similar for all the models. Evaluation of the models' performance on a single class at a time gave us a more clear picture of how the models actually perform. Both YamNet and AlexNet had a relatively stable performance, while ResNet-50 showed evidence of being biased towards certain classes.

None of the models was able to match the natural language query to the corresponding interval with a reliable degree of accuracy and precision. However, there are a few factors that might have affected the experiment negatively. Firstly, YamNet, the best performing model, could have performed better on the classes that had more examples in the original training dataset because had been trained originally with more samples. Secondly, the models might have misclassified the audio samples and it is possible that retrieving the intervals for related classes as well and aggregating that into a single array of intervals would have yielded a more reliable result. However, picking the relevant classes for the audio samples is beyond the scope of the study and could be the baseline for future research.

## 6    Conclusions

The experimental study has compared the performance of three models Yam-Net, AlexNet and ResNet-50 on two different but related problems: a classification problem and interval retrieval based on natural language query. The result was evaluated using the following metrics: ROC AUC, MCC, accuracy, precision and F-1 score. The motivation behind this study was to discover which classifiers are able to solve this problem and to what degree (performance-wise) the classifiers are able to retrieve intervals based on a natural language query.

We used transfer learning for all the three models evaluated by training instances of Random Forest, each of which used the pretrained high level features of the models. The original dataset contains samples that are labelled with 527 classes, however, we used only 6 classes during the experiment to compensate for the lack of the computational capacity and to be able to analyse the result more clearly.

The study established that the tested models were not capable of retrieving intervals from an audio of an arbitrary length based on a natural language query; however, the degree to which the models are able to retrieve the intervals varies depending on the queried keyword and other underlying learning parameters, such as the value of the threshold that is used to filter the audio patches that yield too low probability of the queried class.

The results showed that YamNet was able to classify single fixed-size audio samples with 92.7% accuracy and 68.75% precision, while its average accuracy on intervals retrieval was 71.62% and precision was 41.95%. AlexNet classified the single fixed-size audio samples with 88.7% accuracy and 61.74% precision, while its average accuracy on intervals retrieval was 67.72% and precision was 33.83%. ResNet-50 classified the single fixed-size audio samples with 88.85% accuracy and 63.11% precision. Its average accuracy on intervals retrieval was 71.93% and precision was 41.16%. ResNet-50 yielded an unreliable result on two of the tested classes and displayed a significant difference in the average activation strength of different classes.

# Acknowledgement

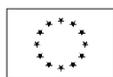This project has received funding from the European Union Horizon 2020 research and innovation programme under grant agreement no. 833673 and no. 786698. This work reflects authors view and Agency is not responsible for any use that may be made of the information it contains.